# Estimation of the New Agile XP Process Model for Medium-Scale Projects Using Industrial Case Studies

M. Rizwan Jameel Qureshi

*Abstract*—Agile is one of the terms with which software professionals are quite familiar. Agile models promote fast development to develop high quality software. XP process model is one of the most widely used and most documented agile models. XP model is meant for small-scale projects. Since XP model is a good model, therefore there is need of its extension for the development of medium and large-scale projects. XP model has certain drawbacks such as weak documentation and poor performance while adapting it for the development of medium and large-scale projects having large teams. A new XP model is proposed in this paper to cater the needs of software development companies for medium-scale projects having large teams. This research may prove to be step forward for adaptation of the proposed new XP model for the development of large-scale projects. Two independent industrial case studies are conducted to validate the proposed new XP model handling for small and medium scale software projects, one case study for each type of project.

## I. INTRODUCTION

All the agile process models emphasis on the need of quality design [1]. Agile models focus on delivering first increment in couple of weeks and complete SW in couple of months. The objective of agile principles is to provide support for the development of only simple and small software projects having small teams [2]. There is no guidance how to customize agile process models for the development of medium and large-scale software projects having large teams. An inadequate amount of experimentation has been carried out to validate agile models for the development of medium and scale projects having large teams [3]. Main characteristics of all agile models are fast development and cost saving. Fast development leads to poor quality SW and carrying all disadvantages of RAD and Prototype process models. Various attempts have been made from last few years to improve agile models [4]. The people who proposed agile software development have provided many facts in support of agile procedures in variety of industrial domains [5]. Most of the projects, using XP model, are validated by using controlled case studies having small teams of four to six members for the implementation/adaptation of existing of XP model on medium and large-scale software projects. Therefore, there is a need for adaptation and validation of XP model for medium and large projects by conducting industrial case studies having large teams. Following is the research question focusing only on medium projects.

How to propose a new XP process model by modifying existing agile XP model for development of medium projects having large teams?

The further paper is organized as following. Section II describes the research design. Section III proposes a new XP model. Section IV presents validation of the proposed new XP model using two case studies.

## II. RESEARCH DESIGN

The design used in this research is case study as research method employed. Two case studies are conducted to validate this research. It may be mentioned that normally results of this type of case studies research is based on 2 to 6 releases. The author deemed it appropriate to conclude the results over here on the basis of 4 releases. The development tool was Microsoft ASP.NET and database tool was Microsoft SQL server for the two case studies conducted for the research problem. The web server was Microsoft Internet Information Server (IIS). The software tools used for the development of the two case projects are shown in Table I.

TABLE I: SOFTWARE TOOLS USED FOR DEVELOPMENT OF THE TWO CASE STUDIES

| Software Tools | Details |
|---|---|
| Operating System | Microsoft XP 2003 |
| Framework | Visual Studio.Net |
| Programming tool | ASP.NET |
| Database tool | Microsoft SQL server |
| Software Configuration Management Tool | Visual SourceSafe (VSS) |
| Unit Testing | NUnit |
| Integration & System Testing | FitNesse |
| Web Server | Internet Information Server (IIS) |
| Documentation Tool | Microsoft Word 2003 |

## III. THE PROPOSED NEW XP METHODOLOGY

The main phases of existing XP model are planning, design, coding and testing [1]. The main phases of the new XP model are 'Project Planning', 'Analysis and Risk Management', 'Design & Development' and 'Testing'. The The author was with COMSATS Institute of Information Technology, Lahore, Pakistan. He is now with the department of Information Technology, Faculty of Computing & Information Technology, King Abdulaziz University, Jeddah, Saudi Arabia (e-mail: anriz@ hotmail.com).

proposed new XP process model is shown in the Fig. 1. The proposed XP process model phases have following activities.

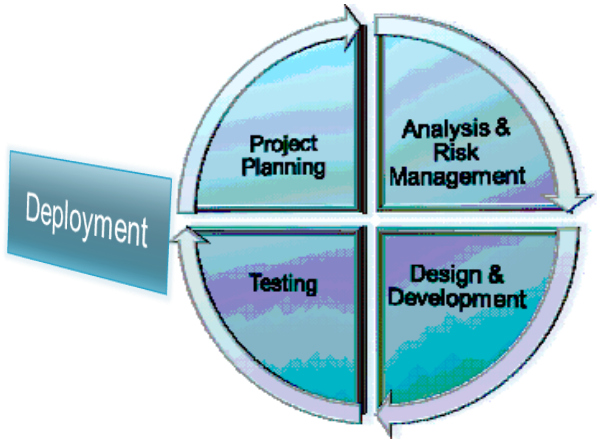

Fig. 1. The Proposed New XP process model.

**Project Planning Phase**- Project specification or proposal document is prepared during the 'Project Planning' phase by communicating it to the customer. Project specification or proposal document is composed of feasibility report that is prepared for cost benefits analysis (CBA). Feasibility report consists of economic, technical and operational feasibilities. Organizational feasibility is prepared based on client request or on project requirements. CBA sheet helps to estimate that software project is feasible for the customer or not. Project team members are also selected during the planning phase. Project team size depends on the size and schedule of the project.

**'Analysis & Risk Management' Phase**- This phase is only started if a customer approves the proposal. Analysis phase improves quality of software through proper documentation. 'Analysis and Risk Management' phase in the proposed XP model has many benefits e.g., initial risk management plan to cater the potential risks regarding the failure of project. There were so many projects failed because of lack of risk management when existing XP model was proposed initially in 2001 [1]. This is the phase where an analyst gathers detailed requirements/user stories. Analysis phase results in more comprehensive user requirements/user stories, modeling and documentation. Comprehensive analysis always gives better design which results in high quality software. Client is requested to prioritize the user stories on his need basis and provides an index value like existing XP model. High index value indicates high priority. Client can: 1) change the order of user stories at any time, 2) provide new user stories at any time. Planning poker technique is implemented to achieve a manageable value (in development weeks) of user stories to be completed in a release. Modeling of the user stories is kept simple to avoid complicacies for the software engineering team. Project velocity is measured after successful deployment of first release and then in all subsequent releases throughout the development of project. The objective is to adjust the delivery dates of remaining user stories to be developed.

**Design and Development Phase**- Design and coding phases of existing XP model are merged to incorporate agility in the new XP process model. The proposed XP process model uses prototype approach to verify the design and user stories. Software is developed in small releases/increments as the customer approves prototypes. Merging of design and development phases also improves efficiency of the software development. Refactoring technique is implemented during designing and coding of a release. Interface specification document is designed for the user stories to be developed in first release. The task of coding is assigned to programmers following pair programming. Interface specification document of user stories of the second release is designed simultaneously by the time programmers coded the user stories of first release. This process of designing & coding is cyclic for the remaining releases till the whole software is developed. The programmers continuously integrated the code (as they completed it for a particular release) into software using Visual Source Safe.

**Testing Phase**-Test cases are prepared for before coding of a release following test-driven development (TDD) environment. Each release is tested on unit basis. Integration test is then conducted to check integration among modules. System test is the next phase to validate the whole increment as one unit. Acceptance testing is the last test to verify a release from the customer. Tested release is maintained and deployed. The proposed new XP process model is cyclic and evolutionary till software development is completed. Main activities of deployment are installation, training and security.

## IV. VALIDATION OF THE PROPOSED NEW XP MODEL

The proposed new XP model is initially validated using two case studies to deal with the research problem for small and medium scale projects. Future work of this research is to adapt and validate for a large-scale project. The results of the two case studies are presented as follows.

**Case study 1** is designed to implement new XP model for a small-scale project. It is conducted for a software company to develop a Logistic Information System (LIS). The results of case study 1 are reported based on the completion of the first four releases of LIS project. Table II shows the data obtained from the four releases for the case study 1.

TABLE II: THE RESULTS OF CASE STUDY 1

| Data Elements | Release 1 | Release 2 | Release 3 | Release 4 |
|---|---|---|---|---|
| No of user stories completed | 5 | 7 | 7 | 6 |
| Line of code/release | 4000 | 6000 | 6000 | 5000 |
| Time to complete (weeks) | 1 | 1 | 1 | 1 |
| Average Productivity (LOC/Hour) | 200 | 225 | 230 | 210 |
| Error Raised/KLOC | 6 | 6 | 5 | 5 |
| Defects/KLOC | 1 | 2 | 1 | 1 |
| Average Maintenance Time (hours) | 2 to 3 | 2 to 3 | 2 to 3 | 2 to 3 |

Data of a case study project for a comparison purpose, about existing XP model for its implementation on a small-scale project, has been taken from the research of the author [6]. Four releases of existing XP project data for a small-scale project are analyzed with new XP project data to conclude the results in Table III using defects rate per Kilo Line of Code (KLOC). It appears by taking a quick look on Table III that so far as suitability of the new XP model versus existing XP model, for a small-scale project, is concerned; both are almost equally good apart from for $4^{th}$ release. The cause for increased defect rate in the $4^{th}$ release of existing XP project is the less usage of pair programming practice.

TABLE III: COMPARISON TABLE OF DEFECTS RATE PER KLOC FOR SMALL SCALE PROJECTS

| Releases | 1 | 2 | 3 | 4 |
|---|---|---|---|---|
| Existing XP [6] | 2.19 | 2.1 | 2.04 | 8.70 |
| New XP | 1 | 2 | 1 | 1 |

**Case study 2** is designed to implement new XP model for a medium scale project. This case study is conducted for a software company to develop an academy management system for an army cadet college on the premises of software company at Pakistan. The case study 2 results are reported based on the completion of the first four releases of academy management system project. Data of a case study project for a comparison purpose, about existing XP model for its implementation on a medium-scale project, has been taken from the research of the authors [7]. A technical comparison of both case study projects is shown in Table IV.

TABLE IV: THE TECHNICAL COMPARISON OF EXISTING XP AND NEW XP CASE STUDIES

| Items | Existing XP Case Study [7] | Case Study 2 |
|---|---|---|
| Case Study Type | Controlled | Industrial |
| Team Size | 4 | 17 |
| Language | Java | ASP.Net |
| Database | MySQL | Microsoft SQL server |
| Software Configuration Tool | CVS | Visual SourceSafe (VSS) |
| Unit, Integration & System Testing | JUnit | NUnit, Fitnesse |
| Code Coverage | Not Provided | NCover |
| Web Server | Apache Tomcat | Microsoft IIS Server |

Four releases of existing XP project data are analyzed with the proposed new XP project data to conclude the results. Table V shows a comparison of exploratory data for the total of first four releases of existing XP model with the proposed new XP model for medium-scale projects.

Table V clearly shows that the proposed new XP model has framed better results compared to existing XP model when implemented on a medium-scale project having large teams in an industrial project. The results of case study 2 support the validation of the proposed new XP model catering the need of software development companies while adaptation of agile XP model for medium-scale projects.

TABLE V: THE EXPLORATORY DATA FOR EXISTING AND NEW XP CASE STUDIES FOR MEDIUM SCALE PROJECTS

| Collected Data | Existing XP Case Study [7] | Case Study 2 |
|---|---|---|
| No of user stories completed | 27 | 31 |
| Line of code | 6629 | 27000 |
| Time to complete (weeks) | 7 | 8 |
| Average productivity (LOC/Hour) | 64 | 865 |
| Error raised/KLOC | Not Provided | 30 |
| Defects/KLOC | 15 | 11 |
| Average maintenance time (hours) | Not Provided | 18 |

## V. CONCLUSION

This paper supports practice of agile software development by proposing a new XP process model that can be adapted according to the requirements of the software project. Adaptive process model is better than existing agile XP models because it eliminates the limitations of large development teams, poor documentation and medium scale software development. The proposed XP model is validated by conducting two case studies on two independent software development companies. The results of the two case studies show that proposed new XP model can be adapted on small and medium scale projects.

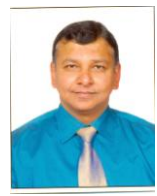

**M. Rizwan Jameel Qureshi** received his Ph. degree from National College of Business Administration & Economics, Pakistan in 2009. He is the best researcher awardee from Faculty of Computing & Information Technology, King Abdulaziz University Saudi Arabia in 2013 and Department of Computer Science, COMSATS Institute of Information Technology Pakistan in 2008.